\documentclass[conference]{IEEEtran}
\IEEEoverridecommandlockouts
\usepackage{cite}
\usepackage{amsmath,amssymb,amsfonts}
\usepackage{algorithmic}
\usepackage{graphicx}
\usepackage{enumerate}
\usepackage{textcomp}
\usepackage{xcolor}
\def\BibTeX{{\rm B\kern-.05em{\sc i\kern-.025em b}\kern-.08em
    T\kern-.1667em\lower.7ex\hbox{E}\kern-.125emX}}

\newcommand{\Lbce}{\mathcal{L}_{\textrm{BCE}}}
\newcommand{\Lmse}{\mathcal{L}_{\textrm{MSE}}}
\usepackage{cite}
\usepackage{url}
\usepackage{array}
\usepackage{comment}
\usepackage{afterpage}
\newcommand{\bhline}[1]{\noalign{\hrule height #1}}   
\newcolumntype{I}{!{\vrule width 1.5pt}}
\begin{document}

\title{LEARNING TO ASSESS SUBJECTIVE IMPRESSIONS FROM SPEECH
\thanks{This
work was supported by JST CREST Grant Number JP-MJCR19A3, Japan.}
}
\author{\IEEEauthorblockN{Yuto Kondo, Hirokazu Kameoka, Kou Tanaka, Takuhiro Kaneko and Noboru Harada}
\IEEEauthorblockA{\textit{NTT Corporation, Japan} \\
\{yuto.kondo, hirokazu.kameoka, kouef.tanaka, takuhiro.kaneko, harada.noboru\}@ntt.com}
}

\maketitle

\begin{abstract}
We tackle a new task of training neural network models that can assess subjective impressions conveyed through speech and assign scores accordingly, inspired by the work on automatic speech quality assessment (SQA). Speech impressions are often described using phrases like `cute voice.' We define such phrases as subjective voice descriptors (SVDs). Focusing on the difference in usage scenarios between the proposed task and automatic SQA, we design a framework capable of accommodating SVDs personalized to each individual, such as `my favorite voice.' In this work, we compiled a dataset containing speech labels derived from both abosolute category ratings (ACR) and comparison category ratings (CCR). 
 As an evaluation metric for assessment performance, we introduce ppref, the accuracy of the predicted score ordering of two samples on CCR test samples. Alongside the conventional model and learning methods based on ACR data, we also investigated RankNet learning using CCR data. We experimentally find that the ppref is moderate even with very limited training data. We also discover the CCR training is superior to the ACR training. These results support the idea that assessment models based on personalized SVDs, which typically must be trained on limited data, can be effectively learned from CCR data. 
\end{abstract}

\section{Introduction}
\vspace*{-0.15cm}
In this paper, we tackle a new task of training neural network models that can assess subjective impressions conveyed through speech and assign scores accordingly. Speech impressions are formed based on para-/non-linguistic information such as the sound and style of speech, often described using phrases like `cute voice,' `youthful voice,' `resonant voice,' and `intelligent voice' in everyday conversation. We refer to such phrases that express the characteristics of the speaker's speech as {\it subjective voice descriptors} (SVDs). The current study draws inspiration from recent research on automatic speech quality assessment (SQA)~\cite{rix2001perceptual,  lo2019mosnet, manocha2022speech, wang2023personalized}, a field that has seen significant attention in recent years. While automatic SQA focuses on evaluating speech solely based on the degradation level of processed speech as perceived by human listeners, our focus lies in scoring any internal impression of speech based on a given SVD.

Conventional audio signal processing has not adequately addressed the needs arising in everyday life. We believe that the proposed task holds great potential for significantly improving the utilization of audio signal processing in everyday scenarios. This potential extends to providing voice training aid for aspiring actors, retrieving voices with specific characteristics from extensive speech databases, and designing voices for text-to-speech synthesis (TTS) systems (Fig.~\ref{fig:example_application}). Furthermore, to further broaden the scope of utilization, it would be advantageous to develop a framework capable of accommodating personalized SVDs, such as `my favorite voice,' which vary from one individual to another.

In line with recent trends in SQA, we find it appropriate to train and test assessment models using both speech and assigned labels. While there are public datasets available for labeling in SQA~\cite{cooper2022generalization}, we are not aware of any datasets specifically tailored for SVDs. Moreover, devising the design and construction of label datasets for SVDs presents a nontrivial challenge. Considering these factors, our objective is to develop a new label dataset pertaining to SVDs. This dataset is based on the following criteria, highlighting the unique disparities between the proposed task and conventional SQA methodologies:
\begin{figure}[t]
 \centering
 \hspace*{-0cm}
\includegraphics[width=0.98\columnwidth]{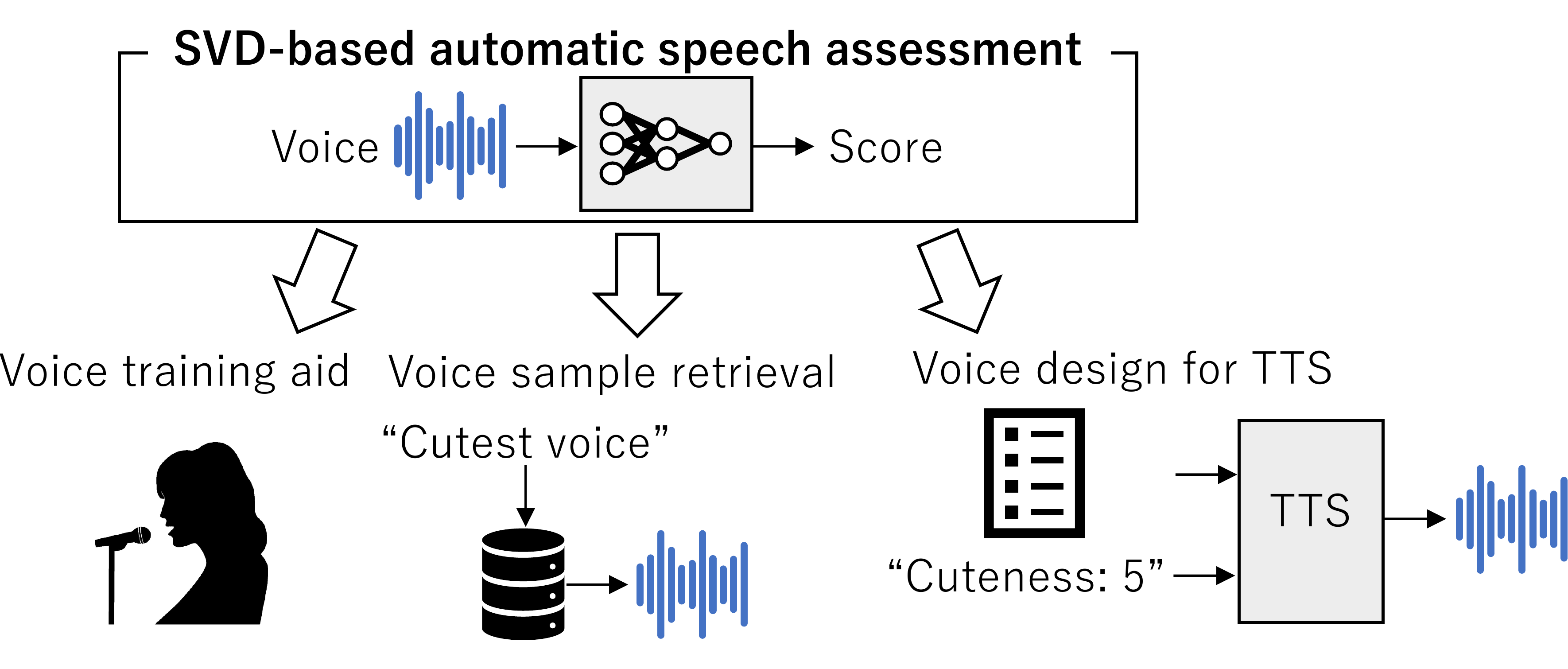}
\vspace*{-0.7cm}
 \caption{A part of utilization of SVD-based automatic speech assessment.}
 
 \label{fig:example_application} 
 \vspace*{-0.68cm}
\end{figure}
\begin{figure}[t]
 \centering
 \hspace*{-0cm}
\includegraphics[width=0.8\columnwidth]{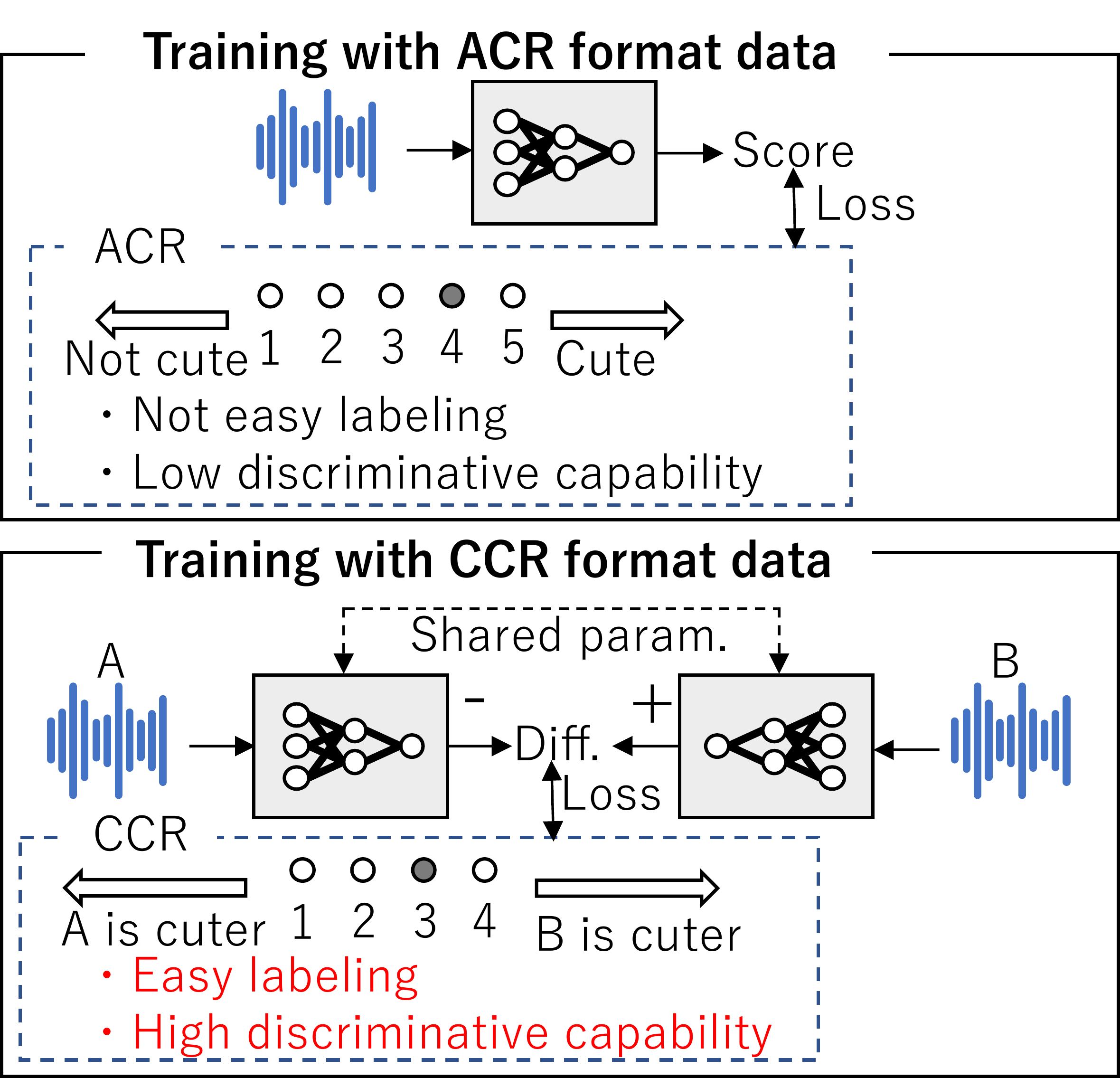}
\vspace*{-0.4cm}
 \caption{Overviews of model training with absolute category rating (ACR) format data, and model training with comparison category rating (CCR) format data. Labeling with CCR format is easier than labeling with ACR format.}
 \label{fig:comp_ACR_pair} 
 \vspace*{-0.7cm}
\end{figure}


\begin{enumerate}[(1)]
    \item As personalized utilization is part of our scope, the framework should have the capability to initially learn solely from data labeled by a single user.
    \item To gather labels from a diverse range of individuals, including children and the elderly, we should employ annotation methods that are user-friendly for beginners. This approach contrasts with SQA labeling, where we could assume high-precision annotations by expert listeners.
    \item We need to utilize a vast number of speakers for annotation due to the diverse range of voice characteristics.
    \item Taking into account the usage scenarios, it is important for assessment results to align with the true order rather than providing an accurate prediction of absolute scores. This contrasts with SQA, where the goal is to predict the absolute score itself, aiming to make the score an objective measure of speech processing systems.
\end{enumerate}
Taking into account the above criteria, we propose devising a label dataset and a development framework that incorporates various refinements not typically found in automatic SQA.
In SQA, assessment models are evaluated using indices based on the absolute category rating (ACR)~\cite{ITU-TP800}, which include metrics like the mean square error (MSE). A common example of ACR is a five-point opinion scale ranging from ``1: bad" to ``5: excellent."
However, when assessing models based on SVDs, we employ the comparison category rating (CCR) ~\cite{ITU-TP800}, which involves random pairwise comparisons, as per the criterion outlined in (4).
Since the CCR is easier to annotate and more reliable compared to the rating of the ACR~\cite{carterette2008here,mantiuk2012comparison}, it meets the criterion outlined in (2).
Furthermore, the CCR may demonstrate greater expressive capability than the ACR because it can capture distinctions between two samples that the ACR cannot discern, potentially resulting in the assignment of the same label.
For example, when considering `cute voice,' all infants' voices might reach the highest point and are difficult to distinguish using the ACR.
This study adheres to the criterion outlined in (3) by not imposing a minimum limit on the number of labels for each sample or sample pair. This approach allows for the inclusion of a diverse range of speech samples from numerous speakers.
This is in contrast to SQA datasets, which provide multiple labels for each sample by limiting the number of speech samples. Considering the criterion outlined in (1), we investigate the model performance under a limited amount of training data. Besides model learning based on ACR training data similar to typical SQA methods, attempts will be made to improve accuracy through CCR training data, leveraging its reliability arising from the ease of scoring (Fig.~\ref{fig:comp_ACR_pair}). Note that such improvements are not necessarily straightforward, as CCR training might not be as efficient as ACR training. This is because $n(n-1)/2$ pair comparisons are needed to encompass the ranking information for $n$ instances, and often only a small portion of all pairs are annotated.

The main contributions of this paper are as follows:
\begin{itemize}
\setlength{\itemsep}{-0cm}
    \item We propose SVD-based automatic speech assessment in order to expand the utilization of speech processing in everyday scenarios.
    \item We design and construct a speech dataset with SVD labels, while being mindful of the disparities between the proposed task and conventional SQA.
    \item Our experimental results demonstrate that models with a reasonable ppref can be learned with a small amount of data, and that CCR learning surpasses ACR learning.
\end{itemize}

The rest of this paper is organized as follows:
In Section \ref{sec:relatedwork}, we discuss related work.
Section \ref{sec:datamake} describes the process of creating the speech dataset with labels based on some SVDs.
In Section \ref{sec:ppref}, we discuss the evaluation metric {\it ppref}.
Section \ref{sec:assessmentmodel} details assessment models and training methods for subsequent experiments.
In Section \ref{sec:exp}, we discuss the experiments.
Finally, we conclude this paper in Section \ref{sec:conc}.

\section{Related work}
\vspace*{-0.15cm}
\label{sec:relatedwork}
\subsection{Automatic speech quality assessment}
\vspace*{-0.15cm}
Extensive research has been conducted on automatic SQA~\cite{rix2001perceptual, lo2019mosnet, manocha2022speech, wang2023personalized} to evaluate speech based on how human listeners perceive the degradation level of processed speech.
One advantage of automatic SQA is that it enables us to evaluate the quality of acoustic equipment or speech processing systems without requiring time-consuming subjective evaluation questionnaires. In recent years, neural automatic SQA methods have been widely investigated to evaluate TTS~\cite{patton2016automos} or voice conversion (VC)~\cite{lo2019mosnet,manocha2022speech, leng2021mbnet,huang2022ldnet} systems. For instance,  MOSNet~\cite{lo2019mosnet} predicts the mean opinion score (MOS) of speech generated by TTS or VC systems, where the MOS is calculated as the average of ACRs.

NORESQA-MOS~\cite{manocha2022speech}, UTMOS~\cite{saeki2022utmos}, MOSPC~\cite{wang2023mospc}, and a preference-based learning method proposed in \cite{hu2023preference} are SQA methods similar to our CCR-based training approach in that they calculate training losses based on the relative relationship between two speech samples. However, while these methods simply compare ACRs (which are relatively less accurate and reliable) as relative ratings, we use CCRs, which are often more reliable, albeit requiring more labor-intensive collection, for relative ratings. 
\subsection{Captioning of sound and style of speech}
\vspace*{-0.15cm}
PromptSpeech~\cite{guo2023prompttts} and Coco-Nut~\cite{watanabe2023coco} are the latest speech corpora along with free-form voice characteristics descriptions. These speech corpora are expected to contribute to development of PromptTTS systems~\cite{guo2023prompttts}, which synthesize speech based on both style and content descriptions. 
Since conventional PromptTTS systems are trained solely on datasets without speech scoring, they may struggle to produce sophisticated expressions that consider the nuanced degree of SVDs.
We believe that scoring based on SVDs make PromptTTS systems more delicate. Additionally, by fine-tuning PromptTTS systems with labels assigned by a single individual using our methodology, personalization of PromptTTS systems is anticipated.

Automatic speaking-style captioning~\cite{yamauchi2023stylecap} is the task of describing speaking styles in speech using natural language. Since it is a captioning task, the potential utilization of this task differs from that of our proposed task.
\begin{figure}[t]
 \centering
 \hspace*{-0cm}
\includegraphics[width=0.7\columnwidth]{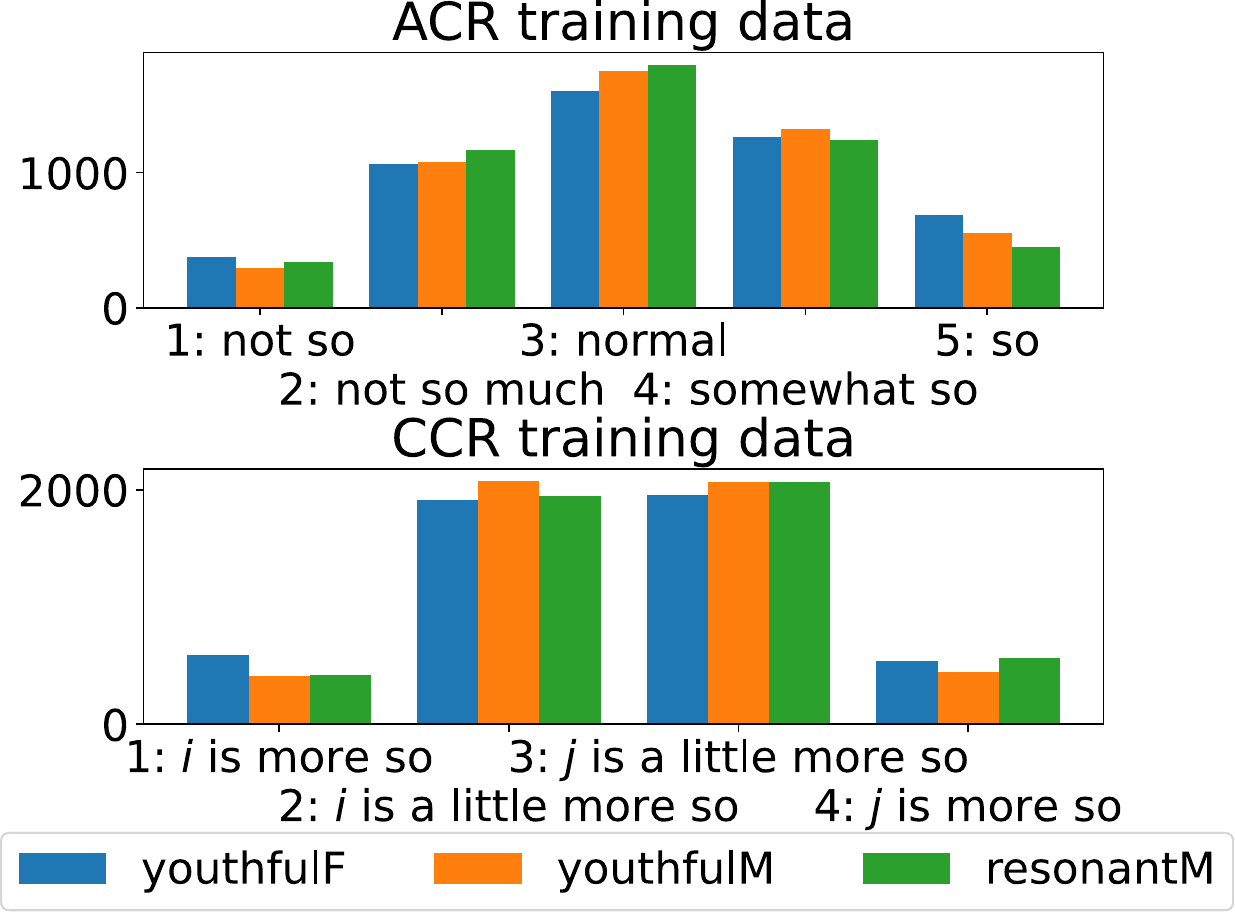}
\hspace*{-0cm}
\vspace*{-0.3cm}
 \caption{Label distributions of ACR and CCR training data.}
 \label{fig:dist_all}
\vspace*{-0.7cm} 
\end{figure}

\section{Data collection}
\vspace*{-0.2cm} 
\label{sec:datamake}
We created the current speech dataset with labels based on certain SVDs through online labeling. The questionnaire, conducted in Japanese, utilized two formats. The first was the ACR format, where participants listened to a single speech sample and assigned a rating from 1 to 5, with options ranging from ``1: not so" to ``5: so." The other format was the CCR, where participants listened to two consecutive speech samples labeled $i$ and $j$ and then chose from four options, indicating whether ``$i$ is more so," ``$i$ is a little more so," ``$j$ is a little more so," or ``$j$ is more so." In the experiment in \cite{mantiuk2012comparison}, forced choice CCR labeling without the option ``no difference" performed better in terms of both accuracy and labeling speed than CCR labeling with the option. Therefore, the option ``no difference" was not included in the CCR options. 

For the speech samples, we utilized a subset of 15 types of Japanese sentences read by $786$ male speakers and $1351$ female speakers from the ATR multi-speaker speech database APPBLA. The average length of all speeches was $5.0$ s. The time-averaged sound level excluding silent intervals was normalized. 
To handle more speech samples, we randomly selected speech samples or pairs of two speech samples to assign labels without setting a lower limit on the number of labels for each sample or pair. This was done on the condition that the two samples comprising each pair are of the same sentence and of the same gender. We randomly selected and played $10$ speech samples as part of the pre-labeling preparation to get a rough sense of overall tendencies of the audio dataset.

While we include personalized utilization in the scope, we used many respondents for the efficiency of label collection, but restricted the use to non-personalized SVDs. The participants comprised $25$ males and $25$ females, all native Japanese speakers. We deliberately selected SVDs that showed relatively little variation in opinion among participants in the ACRs; To summarize broadly, we selected `youthful-sounding voice' of females (youthfulF), `youthful-sounding voice' of males (youthfulM), and `resonant voice' of males (resonantM) among the SVDs extracted from candidates in the literature~\cite{kido1999hyougengo}. These SVDs exhibited relatively large pseudo F~\cite{calinski1974dendrite} values in our preliminary casual experiment. The pseudo F represents the ratio of intra-speaker variance to inter-speaker variance of the scores. 
\section{Evaluation metric for assessment models}
\label{sec:ppref}
\subsection{Precision of preferences~\cite{carterette2008here}}
\vspace*{-0.15cm}
In SQA studies, the MSE, linear correlation coefficient, and Spearman's rank correlation coefficient~\cite{spearman1961proof} are often used as evaluation metrics to measure the discrepancy between predicted and actual scores for ACR-formatted test data.
In light of the usage scenarios of SVD-based speech assessment and the reliability of CCRs over ACRs, we utilize a CCR-based metric, the precision of preferences (ppref)~\cite{carterette2008here}. The ppref is defined as the precision of the predicted score orderings of two samples with respect to the correct binary preference labels. We determine the ordering according to the scores predicted by assessment models. We try two types of pprefs: one is calculated on the binary labels ``$i$ is more so" and ``$j$ is more so" (ppref-strong), and the other is calculated on the binary labels ``$i$ is a little more so" and ``$j$ is a little more so" (ppref-weak).
The ppref-strong is expected to be higher than ppref-weak since the reliability of the labels ``$i$ is more so" and ``$j$ is more so" is higher than that of ``$i$ is a little more so" and ``$j$ is a little more so."
\subsection{Estimation for upper bound of ppref}
\begin{table}[t]
\caption{Estimated upper bounds for ppref-strong and ppref-weak.}
\vspace*{-0.2cm}
\label{tab:est_upper_ppref}
\tabcolsep = 1.5pt
\centering
{
 \begin{tabular}{cIc|c|c} \bhline{1.5pt}
 &\begin{tabular}{c}
Youthful F
\end{tabular} &\begin{tabular}{c}
Youthful M
\end{tabular}&\begin{tabular}{c}
Resonant M
\end{tabular} \\ \bhline{1.5pt}
\begin{tabular}{c}
Ppref-strong
\end{tabular}
&0.922&0.891&0.907\\\hline
\begin{tabular}{c}
Ppref-weak
\end{tabular}
&0.792&0.726&0.756
\\\bhline{1.5pt}
 \end{tabular}
}
\vspace*{-0.5cm}
\end{table}
It is considered that there is an upper bound on ppref, which is less than $1.0$ and varies for each SVD. This is because evaluations for the same sample can vary within and between participants. We estimate one upper bound for ppref-strong and one for ppref-weak for each SVD. Besides the dataset created in Section~\ref{sec:datamake}, we administered $50$ common CCR questions to all participants. 
The speech sample pairs were randomly selected in a manner similar to Section~\ref{sec:datamake}.
We used the average agreement among the participants across all questions as the estimated value for the upper bound of ppref. Here, the agreement was calculated as $\max(a_1,a_4)/(a_1+a_4)$ and $\max(a_2,a_3)/(a_2+a_3)$ for ppref-strong and ppref-weak, respectively, where $a_1$, $a_2$, $a_3$, and $a_4$ represent the number of responses for  ``$i$ is more so," ``$i$ is a little more so," ``$j$ is a little more so," and ``$j$ is more so," respectively.
We present the estimated upper bounds in Table~\ref{tab:est_upper_ppref}.
\section{Assessment models}
\vspace*{-0.15cm}
\label{sec:assessmentmodel}
\subsection{Base models}
\vspace*{-0.15cm}
\label{sec:methods}
We utilized two model architectures for assessment based on SVDs: One is the same as MOSNet~\cite{lo2019mosnet}, which predicts an utterance-level score from the $257$-dimensional magnitude spectrogram using convolutional neural networks (CNNs), followed by a bidirectional long short-term memory (BLSTM) layer, and then a fully connected (FC) layer. We refer to the first architecture as CNN-BLSTM-FC. The other architecture predicts an utterance-level score from the time-averaged $768$-dimensional wav2vec2.0~\cite{baevski2020wav2vec} feature of the input speech waveform through two FC layers, where the wav2vec2.0 is a self-supervised learning model frequently used in some SQA methods, including SSL-MOS~\cite{cooper2022generalization} and UTMOS~\cite{saeki2022utmos}. We use the base wav2vec2.0 model, wav2vec\_small.pt, pretrained on LibriSpeech~\cite{panayotov2015librispeech}. The checkpoint is included in the fairseq toolkit~\cite{ott2019fairseq}. The dimension of the output of the first FC layer is set to $256$, with ReLU and dropout applied after the FC layer. Since we only train the parameters of the FC layers, we call the second architecture fixedSSL-FC.
\subsection{Training methods}
\vspace*{-0.15cm}
We tried two types of training methods: one is the ACR approach, similar to basic SQA methods, and the other is the proposed CCR approach.

In the ACR-based approach, models are trained using backpropagation based on the MSE $\Lmse=(f_{\theta}(x_i)-m_{i})^2$,
where $\theta$ is the set of model parameters to be trained, $f_{\theta}$ is the prediction model, $i$ is the speech index, $x_i$ is the input magnitude spectrogram or SSL feature, and $m_{i}$ is the five-point ACR. 

On the other hand, in the CCR-based approach, models are trained using RankNet~\cite{burges2005learning}. The model parameters are updated by backpropagation based on binary cross-entropy loss: $\Lbce=P_{i, j}\log \hat{P}_{i, j}+(1-P_{i, j})\log(1-\hat{P}_{i, j})$,
where $i$ and $j$ are speech indices, $P_{i, j}$ is the true probability that $j$ is perceived better than $i$ in the context of RankNet, and $\hat{P}_{i, j}$ is the sigmoid-processed difference of predicted scores:
\vspace*{-0.1cm}
\begin{align}
    \hat{P}_{i, j}=\frac{1}{1+e^{-(f_{\theta}(x_j)-f_{\theta}(x_i))}},
\end{align}
\vspace*{-0.1cm}
where $x_i$ and $x_j$ represent the spectrograms of speech $i$ and $j$, respectively, in the case of CNN-BLSTM-FC, or the wav2vec2.0 features in the case of FixedSSL-FC.
In this paper, $P_{i, j}$ is set to $0$, $0.25$, $0.75$, and $1$ corresponding to the assigned CCR labels ``$i$ is more so," ``$i$ is a little more so," ``$j$ is a little more so," and ``$j$ is more so," respectively. RankNet is utilized to train models to produce scores that align with the relative relationships represented by the assigned CCR labels. Consistency in this regard is crucial for the effective utilization of SVD-based assessment models.

\vspace*{-0.05cm}
\section{Experiment}
\vspace*{-0.15cm}
\label{sec:exp}
In this section, we experimentally examine that models with a reasonable ppref can be learned with a small amount of data, and that CCR learning surpasses ACR learning. We successively discuss the experimental conditions and the results of ppref. 
\vspace*{-0.15cm}
\subsection{Experimental conditions}
\vspace*{-0.15cm}
\label{sec:exp_condition}

As described in Section~\ref{sec:datamake}, we used the SVDs youthfulF, youthfulM, and resonantM.
As for the assessment model architecture, we utilized CNN-BLSTM-FC and fixedSSL-FC explained in Section~\ref{sec:methods}.

The dataset was divided as follows to ensure that at least one of the two speakers for each CCR test label was not used as training data:
The training speaker set $S_{\rm train}$ was constructed by randomly selecting from all speakers, comprising $870$ female individuals for youthfulF, and $500$ male individuals for youthfulM and resonantM, respectively. $5000$ ACRs and $5000$ CCRs randomly selected from all the data only labeled to the speakers included in $S_{\rm train}$ were used as the training data. Figure~\ref{fig:dist_all} shows the label distributions of the training data. The test data are a portion of CCRs assigned to the speaker pair including at least one speaker not included in $S_{\rm train}$. For each SVD, we tried $125$, $250$, $500$, $1000$, $2000$, $4000$, and $5000$ as the amounts of training data. The amount of test data used for calculating ppref-strong and ppref-weak was set to $1450$ for both.

The sampling frequency was $16$~kHz. For short-time Fourier transform, Hamming window was used as a window function. The window length and the shift length were $32$~ms and $16$~ms, respectively. The dropout rate was $0.3$. The network parameters except for wav2vec2.0 parameters were initialized by use of Xavier normal initialization. We tried five different initial values. The model was trained using the Adam optimizer with a learning rate of $0.0001$. The batch size was six. The training was run for $30$ epochs.

\subsection{Ppref results}
\vspace*{-0.15cm}
\label{sec:exp_result}
\begin{figure}[t]
 \centering
 \hspace*{-0cm}
\includegraphics[width=1.0\columnwidth]{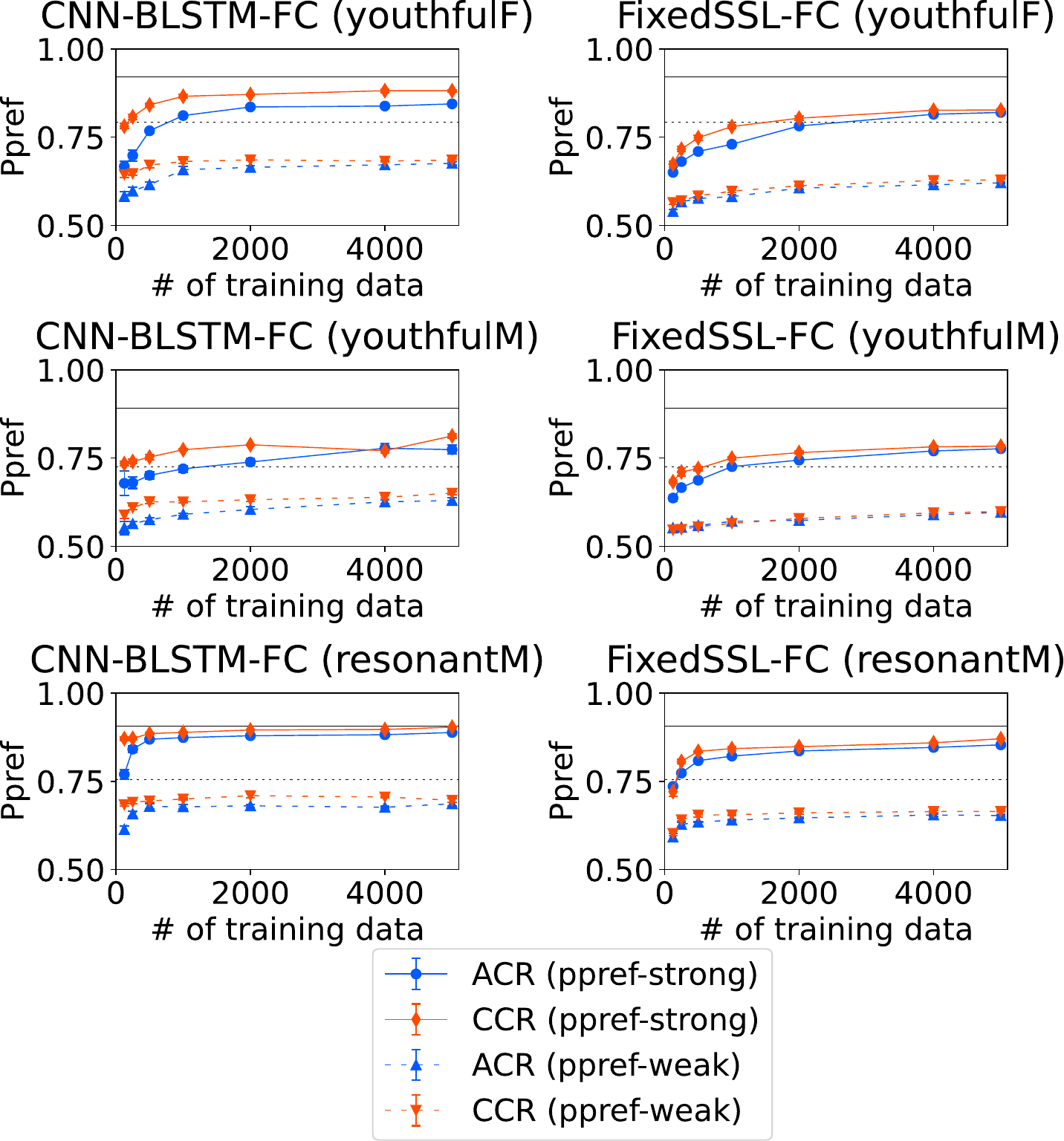}
\hspace*{-0cm}
\vspace*{-0.7cm}
 \caption{Averaged values for five trials of maximum pprefs over all epochs. Horizontal solid and dashed lines represent estimated upper bounds of ppref-strong and ppref-weak, respectively. Estimated upper bounds are shown in Table \ref{tab:est_upper_ppref}.}
 \label{fig:pprefs}
\vspace*{-0.7cm} 
\end{figure}

Figure \ref{fig:pprefs} shows the averaged values across all trials of maximum pprefs over all epochs. The error bars represent standard deviations.
While the ppref-strong generally improves with more training data, the ppref-strong is moderate even with very limited training data. This result encourages the learnability of the assessment models based on personalized SVDs, which are trained with very limited amount of labels obtained from a single user.

Overall, the ppref-strong tends to be better by CCR training than ACR training in all the SVDs and models. 
Though the CCRs used for training are a limited subset randomly extracted from a large pool of pair candidates, they have proven to be valuable training data for SVD-based assessment models. This is considered to stem from the reliability of the CCRs over the ACRs. The superiority of the CCR training over the ACR training is particularly pronounced in CNN-BLSTM-FC when the amount of training data is very limited. Therefore, it is believed that utilizing CCR training data is a favorable option when training assessment models based on personalized SVDs, which are expected to be trained with limited data.

The ppref-strong is higher than the ppref-weak in all the SVDs and methods. Moreover, even with an increase in the amount of training data or the utilization of CCR training data, there tends to be a very limited increase in ppref-weak, despite substantial growth in ppref-strong. The options ``$i$ is a little more so" and ``$j$ is a little more so" are assigned with less confidence compared to the options ``$i$ is more so" and ``$j$ is more so", which would lead to less reliability of the ppref-weak. In future research, it is believed to be beneficial to utilize the ppref-strong for performance evaluation after labeling with the same four-choice options as in this study. This corresponds to screening out test data with low confidence.

The gap between the averaged ppref and the upper bound estimated in Section~\ref{sec:ppref} varies significantly between SVDs. This may be attributed to the differences in the difficulty of capturing key features by assessment models between SVDs.
\vspace*{-0.2cm}
\section{Conclusion}
\vspace*{-0.2cm}
\label{sec:conc}
In this paper, we focused on training neural network models capable of assigning scores based on provided SVDs.
To expand the range of applications, we designed and developed a framework capable of incorporating personalized SVDs, considering the distinctions in usage scenarios and data characteristics between our proposed task and conventional SQA methodologies.
To elaborate, initially, we compiled a dataset comprising both ACRs and CCRs. Subsequently, we introduced ppref, which represents the accuracy of predicted score ordering for two samples in CCR test samples, as an evaluation metric for assessment models.
To enhance assessment performance, we employed RankNet learning with CCRs alongside conventional learning using ACRs. 
Through experimentation, we observed moderate ppref even with minimal training data. 
Interestingly, we discovered that CCR training outperformed ACR training, despite being a smaller subset randomly chosen from a larger pool of pairs, suggesting the superior reliability of CCRs over ACRs. 
These findings endorse the notion that assessment models relying on personalized SVDs can effectively learn from CCR data, even when training data is limited.

Our future research plans include constructing assessment models based on personalized SVDs, estimating the upper limit of ppref for various SVDs, and the application of assessment models to TTS systems. We believe that the methodology proposed in this study, including strategies for data collection, evaluation methods for assessment models, and the model training with CCRs, will support future research on SVD-based speech assessment.


\vspace*{-0.25cm} 
\bibliographystyle{IEEEbib}
\bibliography{refs_kondo}

\end{document}